\documentclass[acmsmall,screen,nonacm,authorversion]{acmart}

\AtBeginDocument{%
  }

\setcopyright{none}
\copyrightyear{2024}
\acmYear{2024}
\acmDOI{XXXXXXX.XXXXXXX}

\acmJournal{TOSEM}
\acmVolume{1}
\acmNumber{1}
\acmArticle{1}
\acmMonth{1}

\usepackage{hyperref}
\usepackage[capitalize]{cleveref}
\makeatletter
\newcommand{\crefnames}[3]{%
  \@for\next:=#1\do{%
    \expandafter\crefname\expandafter{\next}{#2}{#3}%
  }%
}
\makeatother
\crefformat{section}{\S#2#1#3}
\usepackage[frozencache,cachedir=.]{minted}
\colorlet{hlcolor}{yellow!70}
\setminted{
  escapeinside=``,
  highlightcolor=hlcolor,
  numbersep=5pt,
  xleftmargin=1em, % hack to push line numbers out of the margin.
  fontsize=\small,
}
\newcommand{\py}[1]{\mintinline[fontsize=\small]{python3}{#1}}

\newcommand{\sql}[1]{\mintinline[fontsize=\small]{sql}{#1}}
\newcommand{\code}[1]{\mintinline[fontsize=\small]{text}{#1}}
\newcommand{\highlight}[1]{%
  {\setlength{\fboxsep}{0pt}\colorbox{hlcolor}{#1}}}
\usepackage{semantic}
\setnamespace{0em}
\setpremisesspace{1em}
\setpremisesend{0.5em}
\usepackage{mathpartir}
\usepackage[colorinlistoftodos,prependcaption,textsize=tiny]{todonotes}

\usepackage{xspace}
\newcommand{\eg}{\emph{e.g.,}\xspace}
\newcommand{\ie}{\emph{i.e.,}\xspace}

\usepackage{siunitx}

\newcommand{\lang}[1]{\mathcal{L}(#1)}

\newcommand{\many}[1]{\bar{#1}}

\makeatletter
\newcommand{\@kw}[1]{\mathbf{#1}}
\newcommand{\cUnit}{\@kw{unit}}
\newcommand{\cInt}{\@kw{Int}}
\newcommand{\cBool}{\@kw{Bool}}
\newcommand{\cString}{\@kw{String}}

\newcommand{\cTrue}{\@kw{true}}
\newcommand{\cFalse}{\@kw{false}}
\newcommand{\cIf}{\mathop{\@kw{if}}}
\newcommand{\cThen}{\mathbin{\@kw{then}}}
\newcommand{\cElse}{\mathbin{\@kw{else}}}
\newcommand{\cVar}{\mathop{\@kw{var}}}
\newcommand{\cWhile}{\mathop{\@kw{while}}}
\newcommand{\cReturn}{\mathop{\@kw{return}}}
\newcommand{\cAssert}{\mathop{\@kw{assert}}}
\newcommand{\cCall}{\mathop{\@kw{call}}}
\newcommand{\cDef}{\mathop{\@kw{def}}}
\newcommand{\cMethod}{\mathop{\@kw{method}}}
\newcommand{\cPre}{\mathop{\@kw{requires}}}
\newcommand{\cPost}{\mathop{\@kw{ensures}}}
\newcommand{\cLang}{\mathop{\@kw{lang}}}
\makeatother
\newcommand{\sig}[1]{\mathit{\textcolor{teal}{sig}}(#1)}
\newcommand{\pre}[1]{\mathit{\textcolor{teal}{pre}}(#1)}
\newcommand{\post}[1]{\mathit{\textcolor{teal}{post}}(#1)}
\newcommand{\norm}[1]{\mathit{\textcolor{teal}{norm}}(#1)}
\newcommand{\checkTypeFun}{\mathit{\textcolor{teal}{extract}}}
\newcommand{\checkType}[1]{\checkTypeFun(#1)}
\newcommand{\fresh}[1]{#1 \mathop{\text{\textcolor{teal}{fresh}}}}
\newcommand{\transTo}{\mathbin{\textcolor{orange}{\hookrightarrow}}}

\begin{document}

\title{FLAT: Formal Languages as Types}
\subtitle{And Their Applications in Testing}

\author{Fengmin Zhu}
\email{fengmin.zhu@cispa.de}
\orcid{0000-0003-4219-0837}
\affiliation{%
  \institution{CISPA Helmholtz Center for Information Security}
  \city{Saarbr\"{u}cken}
  \state{Saarland}
  \country{Germany}
  \postcode{66123}
}

\author{Andreas Zeller}
\email{zeller@cispa.de}
\orcid{0000-0003-4719-8803}
\affiliation{%
  \institution{CISPA Helmholtz Center for Information Security}
  \city{Saarbr\"{u}cken}
  \state{Saarland}
  \country{Germany}
  \postcode{66123}
}


\begin{abstract}
  Programmers regularly use strings to encode many types of data, such as Unix file paths, URLs, and email addresses, that are conceptually different.
However, existing mainstream programming languages use a unified string type to represent them all.
As a result, their type systems will keep quiet when a function requiring an email address is instead fed an HTML text, which may cause unexceptional failures or vulnerabilities.

To let the type system distinguish such conceptually different string types, in this paper, we propose to regard \emph{formal languages as types} (FLAT), thereby restricting the set of valid strings by context-free grammars and semantic constraints if needed.
To this end, email addresses and HTML text are treated as different types.
We realize this idea in Python as a testing framework FLAT-PY.
It contains user annotations, all directly attached to the user's code, to (1) define such \emph{language types}, (2) specify pre-/post-conditions serving as \emph{semantic oracles} or contracts for functions, and (3) fuzz functions via random string inputs generated from a \emph{language-based fuzzer}.
From these annotations, FLAY-PY \emph{automatically} checks type correctness at runtime via \emph{code instrumentation}, and reports any detected type error as soon as possible, preventing bugs from flowing deeply into other parts of the code.
Case studies on real Python code fragments show that FLAT-PY is enable to catch logical bugs from random inputs, requiring a reasonable amount of user annotations.

\end{abstract}





\maketitle

%%
%% Sections
\section{Introduction}\label{sec:introduction}

Processing structured data is a common task in developing various kinds of software.
File system APIs usually takes file paths as inputs to locate and operate the corresponding resources.
A network library must correctly validate and extract data fields such as a domain name from a URL or an email address, according to their RFC standards.
A server backend typically encodes and decodes packets that are transmitted in standard data exchange formats like XML and JSON.

Ideally, values that represent file paths, URLs, email addresses, etc. should be given \emph{different} types simply because they are conceptually different, for instance, by creating different class types in an object-oriented language or different algebraic data types in a functional language.
However, in reality, programmers use a \emph{unified} string type to rule them all, which eases the design of APIs and the exchange of data, particularly in Unix, where file paths, shell commands, and standard input/output streams are text-based.

However, this golden practice ignores the latent structure of strings and can introduce potential bugs.
Consider a Python function that sends a verification email to a user registered on a web service:
\begin{minted}{python3}
def send_verif_email(email: str) -> None
\end{minted}
A static type checker (\eg mypy\footnote{\url{https://mypy-lang.org}} and pyre\footnote{\url{https://pyre-check.org}}) happily accepts any string as the input argument \py{email}, including an XML string and an HTML string injected with malicious JavaScript code.
Later, those undesired inputs can lead to unpredictable behaviors and even security issues that are hard to detect and repair.

One possible industrial solution is to perform additional input format validation through \emph{regular expression} (regex) matching, supported in many mainstream programming languages like Python's \py{re} library.
To ensure that the input \py{email} address is valid in \py{send_verif_email}, we may attach the following validation code at the beginning of this function:
\begin{minted}[linenos]{python3}
import re
def send_verif_email(email: str) -> None:
  pattern = re.compile(r'(?<=[^0-9a-zA-Z\!\#\$\%\&\'\*\+\-\/\=\?\^\_\`\{\|\}\~\-])' \
    r'([a-zA-Z0-9_.-]+@[a-zA-Z0-9_.-]+(?:\.[a-zA-Z0-9]+)*\.[a-zA-Z0-9]{2,6})' \
    r'(?=[^0-9a-zA-Z\!\#\$\%\&\'\*\+\-\/\=\?\^\_\`\{\|\}\~\-])') # regex pattern
  assert pattern.fullmatch(email) is not None # ensure it matches the input email
  ... # now can safely send email
\end{minted}

The regex pattern shown in Lines 2-5 is taken from a Python library JioNLP\footnote{
  \url{https://github.com/dongrixinyu/JioNLP/blob/e17dba0/jionlp/rule/rule_pattern.py\#L43}
}.
It is not only \emph{unreadable} but also \emph{wrong}: it both over- and under-approximates the legal email addresses as defined in RFC 5322 \cite{RFC5322} standard of Internet message format.
For example, it will \emph{reject} the legal email address \py{'name/surname@example.com'} (note that \py{'/'} is allowed) and \emph{accept} the illegal email address \py{'a"b(c)d,e:f;g<h>i[j\\k]l@example.com'} (as \py{','} and brackets \py{'[]'} are disallowed)\footnote{
  Both examples are taken from: \url{https://en.wikipedia.org/wiki/Email\_address\#Examples}.
}.
Regex patterns like this are unreadable, unusable, and may be too complex to be correct, thus are not reasonable options for expressing the exact intended structure of strings.
% Here is an overview of the email addresses syntax written in ANTLR-style\footnote{\url{https://www.antlr.org}} grammar:
% \begin{minted}{antlr}
% email: local_part "@" domain;
% local_part: dot_atom | quoted_string | obs_local_part;
% \end{minted}
% An email address consists of a local-part and a domain, combined with a \py{'@'}.
% A simple case of the local-part is \code{dot_atom}:
% \begin{minted}{antlr}
% dot_atom: CFWS? dot_atom_text CFWS?;
% dot_atom_text: atext+ ("." atext+)*;
% atext: ALPHA | DIGIT | "!" | "#" | "$" | "%" | "&" | "'" | "*" | "+" | "-"
%      | "/" | "=" | "?" | "^" | "_" | "`" | "{" | "|" | "}" | "~";
% \end{minted}
% Ignoring comments, a \code{dot_atom} is a dot-separated string allowing the characters defined in the \code{atext} rule: letters (\code{ALPHA}), digits (\code{DIGIT}), and special characters including slashes (\py{'/'}) but excluding commas (\py{','}).
% Thus, \py{email_1} is valid while \py{email_2} is invalid\footnote{
%   The two example email addresses are taken from: \url{https://en.wikipedia.org/wiki/Email\_address\#Examples}.
% }:
% \begin{minted}{python3}
% email_1 =  # valid, '/' allowed
% email_2 = 'a"b(c)d,e:f;g<h>i[j\\k]l@example.com' # invalid, ',' etc. disallowed
% \end{minted}

Such a regex-based input format validation approach has the following drawbacks:
(1) unwieldy validation code (Lines 3-6 shown above) has to be inserted manually, which is labor-intensive and time-consuming;
(2) regex patterns are less expressive and do not offer higher level formal language support as in \emph{context-free grammars} (CFGs); and
(3) no existing \emph{language-based fuzzing} technique (\eg \cite{Input-Algebras,ISLa,Fuzzing-Book}) takes (Python's or JavaScript's) regex patterns as inputs.

To address these three issues, in this paper, we present a more general solution from a type system's angle based on the idea of regarding \emph{formal languages as types} (FLAT).
The formal languages we use are CFGs: they are well-defined, well-understood, and well-supported by language-based fuzzers.
Even for a regular language, a well-structured CFG with reasonable names for nonterminal symbols is more readable than a compact yet mysterious regex pattern.
Not even mention that some languages are indeed not regular, including the RFC-standard email addresses.
A CFG gives rises to a \emph{language type} whose inhabitants are restricted to the set of sentences accepted by that grammar, hence rejected malformed inputs.
For example, the CFG for email addresses as specified in RFC 5322 derives a language type \py{Email}, with which we provide a more precise type signature for \py{send_verif_email}:
\begin{minted}{python3}
def send_verif_email(email: `\highlight{Email}`) -> None
\end{minted}
Feeding an XML string or a JavaScript-injected HTML string will result in type errors.

With such language type annotated in functions and variables, FLAT \emph{automatically} performs \emph{runtime type checking}.
The type checker reports type errors such as ``the input \py{email} string is not an \py{Email} according to the syntax'' right away, which avoids bugs flowing deeply into other code.
No validation code are needed as in the regex-based approach, but language types.
These types serve as documentations for functions, which can increase the readability of the original code.

Of course we need test inputs to start runtime type checking, and they can be user-provided or randomly-generated.
FLAT leverages \emph{language-based test generation} to produce random sentences from the associated grammars of language types and test user-functions as requested.
We offer a convenient \py{fuzz} function to enable random testing:
\begin{minted}{python3}
def fuzz(target: Callable, num_random_inputs: int=1000)
\end{minted}
Using this, we can test \py{send_verif_email} with 1,000 random inputs with a single line of code:
\begin{minted}{python3}
fuzz(send_verif_email)
\end{minted}

FLAT goes beyond input format validation in two dimensions.
First, FLAT allows users to specify pre- and post-conditions, serving as contracts for functions, from which it can automatically detect and report any violation of these semantic constraints.
That is to say, not only input but also output format, as well as input-output relations can be specified as \emph{semantic oracles} and tested.
Second, FLAT allows attaching additional logical constraints over language types so that \emph{context-sensitivity} is considered.
With these two extensions, FLAT is able to test functional correctness and find logical bugs, preferably in an automated manner using our \py{fuzz} function as introduced above.

\paragraph{Contributions}

To sum up, this paper makes the following contributions:
\begin{itemize}
    \item We present FLAT, a general language-agnostic type-directed testing framework for string-manipulating programs using formal languages as types.
    \item We define an abstract core language FLAT-CORE (\cref{sec:core}, \cref{sec:tyck}) to demonstrate FLAT's type system and runtime type checking.
      We rely on \emph{code instrumention} to check if all input arguments, return values, and local variables match the annotated types.
    \item The key idea of FLAT is not tied to a specific programming language (\cref{sec:migrate}).
      We migrate FLAT-CORE to Python and introduce FLAT-PY (\cref{sec:py}), a testing framework for Python with built-in language types for commonly used data formats (\eg email addresses, URLs, JSON) and Python-style annotations (including the \py{fuzz} function).      
    \item We conducted case studies (\cref{sec:eval}) on open-source Python code fragments processing user profiles and file paths.
      Requiring a reasonable amount of user annotations that are intuitive to specify, FLAT-PY identifies logical bugs from randomly generated test inputs.
\end{itemize}

We start in \cref{sec:tour} with an overview of the motivations and ideas behind FLAT using concrete Python code with FLAT-PY annotations.
We discuss related work in \cref{sec:related} and conclude the paper in \cref{sec:conclusion}.
Our tool and data are publicly available: see \cref{sec:conclusion} for the link.

\section{A Tour of FLAT-PY}\label{sec:tour}

To demonstrate how one uses FLAT-PY to test and debug their code, let us consider an ad hoc parser:
\begin{minted}[linenos]{python3}
def get_hostname(url: str) -> str:
  """Extract the hostname part."""
  start = url.find('://') + 3
  end = url.find('/', start)
  host = url[start:end]
  return host
\end{minted}

This function aims to extract a hostname from a given URL.
It first computes the starting and ending positions of the hostname part, represented by the local variables \py{start} and \py{end} respectively.
The \py{start} (Line~3) is the position right after \py{'://'}.
This position is computed by adding the starting index of the pattern with its length three, where the starting index is obtained via a call of the \py{find} method.
The \py{end} (Line~4) is the position of the first \py{'/'} seen afterward, which is obtained via a call of the \py{find} method within a range beginning with \py{start}.
Finally, using these two positions, the hostname part is extracted as a substring of the input \py{url} (Line~5) and returned (Line~6).

Let us now imagine that we wish to save the extracted hostname in a database table named \py{hosts}.
We achieve this by executing a SQL query that is instantiated from the following template, where \code{<host>} is the metavariable we will substitute:
\begin{minted}{sql}
INSERT INTO hosts VALUES ('<host>')
\end{minted}
The function below summarizes our workflow:
\begin{minted}[linenos]{python3}
def save_hostname(url: str, db_cursor: MysqlCursor):
  sql_temp = "INSERT INTO hosts VALUES ('{host}')"
  hostname = get_hostname(url)
  sql_query = sql_temp.format(host=hostname)
  db_cursor.execute(sql_query)
\end{minted}
It uses the \py{format} method to instantiate the template \py{sql_temp} (Line~4), and the \py{execute} method from MySQL library to fire a query (Line~5).
Is this implementation correct and safe?

\subsection{Detecting SQL Injection}

The previous implementation is indeed unsafe and suffers from SQL injection.
To see this, let us consider the following malicious input, which is indeed not a valid URL (note that ``\code{--}'' starts a single line comment in SQL):
\begin{minted}{text}
"https://localhost'); DROP TABLE users --/"
\end{minted}
Feeding it into the \py{get_hostname} function, we obtain the following output:
\begin{minted}{text}
"localhost'); DROP TABLE users --"
\end{minted}
Instantiating the \py{sql_temp} variable in function \py{save_hostname} with it, we end up having a query involving an undesired \sql{DROP TABLE} command:
\begin{minted}{sql}
INSERT INTO hosts VALUES ('localhost'); DROP TABLE users --')
\end{minted}
Since all the above steps are well-typed, and that no runtime error has occurred so far, this malicious query will be executed, leading to the destruction of the \code{users} table in our database.
Only then we will detect this security issue, but it is too late.

From the above example, we learned that Python's type system cannot detect such security issues caused by strings with unexpected formats.
What we really need is a fine-grained type system that is aware of the internal structures of strings.
In our framework, to enable such a feature, the user needs to specify type annotations using our \emph{language types}.

\paragraph*{Solution 1: annotate \py{get_hostname}}

We expect the input \py{url} to be a URL string, and the return value to be a hostname.
Thus, we annotate this function using two built-in language types \py{URL} and \py{Host}\footnote{
  As this function aims to handle simple URLs with ordinary hostnames (say, not IP addresses) and without queries and fragments, here we use a simplified URL syntax rather than the standard one defined in RFC 3986 \cite{RFC3986}.
  For the latter, we provide another built-in type \py{RFC_URL}.
}:
\begin{minted}{python3}
def get_hostname(url: `\highlight{URL}`) -> `\highlight{Host}`:
    ...
\end{minted}

FLAT-PY now processes the above type annotations, and generates an \emph{instrumented} version with two type assertions: one checks if the input \py{url} is a sentence of the URL grammar, and the other checks if the return value is a sentence of the hostname grammar, via parsing.
If either check fails, a type error will be reported.
Feeding the aforementioned malicious input to the instrumented code, we will encounter the following type error:
\begin{minted}{text}
Type mismatch for argument 0 of method get_hostname
  expected type: URL
  actual value:  "https://localhost'); DROP TABLE users --/"
\end{minted}
The error message says that the malicious input is ill-typed, as it does not match the URL grammar (\py{';'} is disallowed).
This error aborts the execution and avoids the SQL injection issue.

\paragraph*{Solution 2: annotate \py{sql_query}}

An alternative approach is to annotate the local variable \py{sql_query} defined in \py{save_hostname} with a type for ``safe'' SQL queries:
\begin{minted}{python3}
sql_query: SafeSQL = sql_temp.format(host=hostname)
\end{minted}
This customized language type \py{SafeSQL} can be introduced using a special type constructor \py{lang} provided by FLAT-PY:
\begin{minted}{python3}
SafeSQL = lang('SafeSQL', """
start: "INSERT INTO hosts VALUES " "(" Host ")";
""")
\end{minted}
The second argument defines a grammar for this language type using an ANTLR-like \cite{ANTLR} notation, and one is allowed to directly refer \py{Host} to the hostname grammar.
Note that this grammar rejects any query that indeed contains a non-insert command such as the dangerous \py{DROP TABLE} command.

So far, we have finished all the required annotations.
FLAT-PY now takes over to generate the instrumented code.
Feeding the aforementioned malicious input to it, a type error will be triggered:
\begin{minted}{text}
Type mismatch
  expect:    SafeSQL
  but found: "INSERT INTO hosts VALUES ('localhost'); DROP TABLE users --')"
\end{minted}
Again, we avoid the SQL injection issue.

\subsection{Language-Based Test Generation}\label{sec:tour:test-gen}

Apart from the security issue, is our implementation functionally correct?
That is to say, does it behave normally when feeding valid inputs?
To answer this question, \emph{language-based fuzzing} is a low-labor-cost approach and is supported by FLAT-PY through the \py{fuzz} function.

To fuzz the \py{get_hostname} function, the user calls:
\begin{minted}{python3}
fuzz(get_hostname, num_random_inputs=50)
\end{minted}
The first argument is the test target and the second is the number of random inputs needed.
Since the sole argument \py{url} of \py{get_hostname} is annotated with the language type \py{URL}, our framework will automatically synthesize a producer for URL strings, built from an existing language-based fuzzer ISLa \cite{ISLa}.
ISLa can produce random sentences not only complying with a CFG, but also statisfying additional logical constraints via its \emph{Input Specification Language}, which is helpful for FLAT-PY to handle context-sensitivity (will see later).

From the 50 random test inputs, a type error caused by empty-path URLs was detected:
\begin{minted}{text}
Type mismatch
  expect:    Host
  but found: ''
\end{minted}
This error message says that a failing input \py{"http://W"} was found: note that it is indeed a valid URL.
However, the output value---the empty string---is not a valid hostname.
But for this particular input, the hostname should be \py{"W"} rather than empty, thus our implementation is incorrect.

Through further debugging, we realized this bug was caused by the code at Line~4: since the character \py{'/'} was not found in the rest part of the input \py{"W"}, the call to \py{find} returned \py{-1}, leading to an incorrect ending index of the hostname part.
Consequently, the function outputs an empty string that violates the \py{Host} type requirement.

To fix this issue, one can, for example, use an if-statement for a case analysis on the emptiness of the path:
if it is empty, then the hostname should be the remainder of the string;
otherwise, the current approach applies.
The fixed version is shown below (added lines are highlighted):
\begin{minted}[highlightlines={5,6}]{python3}
def get_hostname(url: URL) -> Host:
  """Extract the hostname part."""
  start = url.find('://') + 3
  end = url.find('/', start)
  if end == -1:      # fixed code
    end = len(url)   # fixed code
  host = url[start:end]
  return host
\end{minted}

\subsection{Oracles}\label{sec:tour:oracles}

There is one more property we wish to validate: the output must be the exact hostname extracted from the input \py{url}.
This is indeed an invariant connecting the input and the output and can be used as a test \emph{oracle} for functional correctness.
It cannot be directly encoded as a language type but as a post-condition attached by a special decorator \py{ensure} provided by FLAT-PY.

Using Python's string library functions alone is insufficient to express this condition, as one shall extract a substring (namely, the hostname) according to the syntax, while most of the library functions do not consider such syntax or structure.
Instead, FLAT-PY introduces \emph{XPaths} that locate subtrees in the \emph{derivation tree} of a string, and offers a \py{select} function to extract the substring of the subtree located by a given XPath.
Note that this function is realizable because ISLa \cite{ISLa}, the backend fuzzer we rely on, caches the derivation trees for the generated random inputs and allows such XPaths in its input specification language.

Using such a \py{select} function, we attach the needed post-condition:
\begin{minted}{python3}
@ensures(lambda url, ret: ret == select(xpath(URL, "..host"), url))
def get_hostname(url: URL) -> Host:
  ...
\end{minted}
We extract the hostname substring from \py{url} using the XPath \py{"..host"}:
it refers to the (unique) node labeled with \py{host}, a nonterminal symbol of the URL grammar that refers to the hostname part, in the derivation tree of \py{url}.
The last parameter \py{ret} of the lambda expression binds to the return value.
This post-condition will be validated whenever the function returns. 
No more type errors were detected through fuzzing.

\section{FLAT-CORE: The Core Language}\label{sec:core}

We begin the technical presentation of our FLAT framework with an abstract core language called FLAT-CORE.
It is a minimal language that integrates FLAT's language types into a tiny imperative programming language.
It also offers annotations for users to attach pre- and post-conditions for methods.
It can be extended to combine other advanced language features so that one can build FLAT instantiations for existing programming languages.

\paragraph{Program}

A FLAT-CORE program $p$ contains a set of definitions:
$$\begin{array}{rcll}
  p &::=&  \many{d} & \text{(program)} \\
  d &::=&  \cDef f(x_1: t_1, \ldots, x_k: t_k): t = e & \text{(fun def)} \\
    &\mid& \cMethod m(x_1: t_1, \ldots, x_k: t_k): t~\many{\varphi}~\{\many{s}\} & \text{(method def)} \\
    &\mid& \cLang L = G & \text{(lang def)} \\
  \varphi &::=&  \cPre e \mid \cPost e & \text{(pre, post)} \\
\end{array}$$
The meta notation $\many{\cdot}$ refers to repeating $\cdot$ zero or multiple times.
Like in Dafny \cite{Dafny} we distinguish between functions and methods: one uses functions to write specifications that should be pure, and methods to model the verification/testing targets that are usually impure.
Thus, a function body is an expression $e$, while a method body is a block of statements $\{\many{s}\}$.
We require the user to annotate the types $t_i$ for each parameter $x_i$ and the return type $t$.
A method can optionally have pre-/post-conditions specified using $\cPre$ and $\cPost$ clauses.
A pre-condition is a \emph{predicate} (\ie a Boolean function) over the method's input arguments.
A post-condition is a predicate over the method's input arguments and its return value.
Both are typically lambda expressions.
Furthermore, the special $\cLang$-definition specifies a \emph{language type} $L$ via a grammar $G$: any string value $s$ has this type iff $s \in \lang{G}$.

\paragraph{Grammars}

We adopt an ANTLR-like \cite{ANTLR} meta notation for expressing a grammar in the $\cLang$-definition.
A grammar $G$ consists of a set of production rules $\{r_1; r_2; \ldots; r_n\}$.
Each production rule $A \to \alpha$ consists of a nonterminal $A$ and a clause $\alpha$ that is inductively defined as follows:
$$\begin{array}{rcll}
  \alpha &::=&  a \mid A \mid \alpha_1 \alpha_2 \mid (\alpha_1 \mid \alpha_2) & \text{(standard)} \\
         &\mid& \alpha* \mid \alpha+ \mid \alpha? \mid \alpha\{k\} \mid \alpha\{k_1, k_2\} & \text{(repetition)} \\
         &\mid& [c_1 \text{-} c_2] & \text{(char set)}
\end{array}$$
Like ANTLR, we support terminals (a string literal $a$), nonterminals ($A$), concatenation (whitespace), alternatives (``$\mid$''), Kleene star (``$*$''), Kleene plus (``$+$'') and optional (``$?$'').
In addition, we support convenient notations inspired by Python's regex syntax:
\begin{itemize}
  \item $\alpha\{k\}$ for repeating $\alpha$ exactly $k$ ($k \ge 2$) times,
  \item $\alpha\{k_1, k_2\}$ for repeating $\alpha$ at least $k_1$ times and at most $k_2$ times ($k_1 < k_2$), and
  \item $[c_1 \text{-} c_2]$ for the character set in between $c_1$ and $c_2$ (both inclusive).
\end{itemize}

\paragraph{Types}

We adopt a two-layer type system.
A \emph{simple type} constitutes built-in types (for integers, Booleans, and strings) and function types:
$$\begin{array}{rcll}
  \tau &::=&  \cInt \mid \cBool \mid \cString & \text{(built-in type)} \\
    &\mid& (\tau_1, \ldots, \tau_k) \to \tau & \text{(function type)}
\end{array}$$
The full type system supports language types introduced by $\cLang$-definitions and \emph{refinement types}:
$$\begin{array}{rcll}
  t &::=&  \tau & \text{(simple type)} \\
    &\mid& L & \text{(language type)} \\
    &\mid& \{x : t \mid e\} & \text{(refinement type)}
\end{array}$$
A refinement type $\{x : t \mid e\}$ consists of a \emph{base type} $t$ and a $\cBool$-typed expression $e$ as the \emph{refinement}.
We use the standard set-comprehension notation to indicate the inhabitants of a refinement type is a set of $t$-typed values that satisfy the condition $e$, where the value is bound to $x$ in $e$.
For example, a refinement type for a positive integer is written $\{n : \cInt \mid n > 0\}$.
The bound variable $x$ is ignored if not used in the refinement.

To fit the setting of testing, base types cannot contain function types: one usually needs first-order logic propositions (\ie with quantifiers $\forall$ and $\exists$) to express interesting properties over functions, but they cannot be evaluated to a Boolean value at runtime.

On the other hand, refinement types can be nested, say $\{\{n: \cInt \mid n > 0\} \mid n < 10\}$.
One may put the two conditions ``$n > 0$'' and ``$n > 10$'' together to obtain an equivalent but compact form $\{n: \cInt \mid n > 0 \land n < 10\}$.
We call such an operation \emph{type normalization}, where the normalized type is a refinement type whose base type is always a simple type:
\begin{mathpar}
  \norm{\tau} = \{\tau \mid \cTrue\}
  \and \norm{L} = \{s: \cString \mid s \in L\}
  \\ \inference{\norm{t} = \{y : \tau \mid e_1\}}{\norm{\{x : t \mid e\}} = \{x: \tau \mid e_1[x/y] \land e\}}
\end{mathpar}

\paragraph{Expressions}

We consider standard expressions with two special constructs:
%  a built-in operation $e \in L$ for testing if $e$ has the language type $L$:
$$\begin{array}{rcll}
  e &::=&  n \mid x & \text{(constants, variables)} \\
    &\mid& e(e_1, \ldots, e_k) & \text{(function application)} \\
    &\mid& (x_1, \ldots, x_k) \to e & \text{(lambda expressions)} \\
    &\mid& \cIf e \cThen e_1 \cElse e_2 & \text{(if-then-else)} \\
    &\mid& e \in L & \text{(language type test)} \\
    &\mid& e[\pi] & \text{(XPath selection)} \\
\end{array}$$

The special construct $e \in L$ tests (hence it has type $\cBool$) if the $\cString$-typed expression $e$ has the language type $L$.
The other special construct $e[\pi]$ selects a substring from the $\cString$-typed expression $e$ by an \emph{XPath} $\pi$ (will be introduced soon), \ie the \py{select} function shown in \cref{sec:tour:oracles}.

Note that ordinary unary/binary expressions are internally desugared to function application, for example, $x > 0$ is internally just $\textit{gt}(x, 0)$, where $\textit{gt}$ is a built-in function for greater than on integers.
But for better readability, we prefer unary/binary expressions as syntactic sugars.

\paragraph{Library functions and XPaths}

The core language comes with a library including:
(1) standard unary/binary operations (for arithmetic, comparison, and Boolean computation) over integers and Booleans;
(2) string operations defined in SMT-Lib;
and (3) a special function to extract substrings via an XPath.

For ease of accessing substrings of a string $s \in L$, which are indeed the nodes in the derivation tree of $s$, we adopt an XPath-like syntax, using the nonterminals of the grammar as labels.
An XPath is a sequence of \emph{selectors}, each of which is one of the following:
\begin{itemize}
  \item ``$.A[k]$'': to select the $k$-th $(k \ge 1)$ direct child with label $A$;
  \item ``$.A$'': to select all direct children with label $A$;
  \item ``$..A$'': to select all descendants (both direct and indirect children) with label $A$.
\end{itemize}
For example, the XPath ``$.A[1]..B.C$'' refers to the sentences of:
\begin{itemize}
  \item all the direct children with label $C$
  \item of all descendants with label $B$
  \item in the first child with label $A$ of the root tree.
\end{itemize}

\paragraph{Statements}

We consider commonly-used statements in a typical imperative language:
$$\begin{array}{rcll}
  s &::=&  \cVar x: t; \mid x = e; & \text{(decl, assign)} \\
    &\mid& \cVar y = \cCall m(e_1, \ldots, e_k); & \text{(method call)} \\
    &\mid& \cAssert e; & \text{(assertions)} \\
    &\mid& \cReturn e; & \text{(return)} \\
    &\mid& \cIf e~\{\many{s}\} \cElse \{\many{s}\} \mid \cWhile e~\{\many{s}\} & \text{(control flow)}
\end{array}$$

We allow a declare-and-assign statement ``$\cVar x = e;$'' where the type of $x$ is inferred from $e$.
It is conceptually equivalent to ``$\cVar x: t; x = e;$'' given that $e$ has type $t$.

As the core language distinguishes methods from functions, here we introduce a method-call statement to invoke methods.
Since statements can only appear in the method body, one can only call methods from methods, but not functions.

\paragraph{Example}

Using the core language, we rewrite the Python function \py{get_hostname} defined in \cref{sec:tour} as the method below:
$$\begin{array}{l}
  \cMethod \text{getname}(\text{url}: \text{URL}): \text{Host} \\
  \quad  \cPost~(\text{url}, \text{ret}) \to \text{ret} == \text{url}[..\mathit{host}] \\
  \{ \\
  \quad  \cVar~\text{start}: \cInt = \mathit{indexof}(\text{url}, \texttt{"://"}, 0) + 3; \\
  \quad  \cVar~\text{end}: \cInt = \mathit{indexof}(\text{url}, \texttt{"/"}, \text{start}); \\
  \quad  \cVar~\text{host}: \cString = \mathit{substr}(\text{url}, \text{start}, \text{end} - \text{start}); \\
  \quad  \cReturn~\text{host}; \\
  \}
\end{array}$$

This method contains a post-condition that uses the select operator to extract the substring associated with the XPath ``$..\mathit{host}$''.
The method body uses two SMT-LIB functions from the theory of Unicode strings\footnote{
  \url{https://smt-lib.org/theories-UnicodeStrings.shtml}
}:
\begin{itemize}
  \item $\mathit{indexof}(s, t, i)$ is the SMT-LIB function \code{str.indexof}: it returns the index of the first occurrence of the string $t$ in string $s$, starting at the index $i$;
  \item $\mathit{substr}(s, i, n)$ is the SMT-LIB function \code{str.substr}: it evaluates to the longest substring of $s$ of length at most $n$, starting at the index $i$.
\end{itemize}

\section{Runtime Type Checking via Instrumentation}\label{sec:tyck}

After showing the syntax and semantics of FLAT-CORE, this section presents how to ``type check'' a FLAT-CORE program.
Since this paper focuses on the dynamic approach, the ``type checking'' we refer to here is not the traditional static type checking done at compile time, but \emph{instrumenting} necessary type assertions, each checks if a concrete value has the specified type.
This instrumentation process is done \emph{automatically} via a static analysis of the type definitions and the annotations in the original program.
In the end, we execute the instrumented version: any violated assertion will immediately raise a (runtime) type error.

\subsection{Preparation}\label{sec:tyck:static}

There are trivial type errors in FLAT-CORE programs that are related to simply types merely (not related to any logical constraint), such as assigning an integer variable with a string value, or feeding a Boolean to a function that expects a string.
One can easily identify them via a standard static type checking algorithm (as done in Java), ignoring all the syntactic constraints required by language types (\ie regarding them as the $\cString$ type) and all the semantic constraints occurred in refinement types and pre-/post-conditions.
If the program is free of trivial errors, then we will perform program instrumentation to identify nontrivial errors related to the violation of any syntactic or semantic constraint.

\subsection{Instrumentation}\label{sec:tyck:dynamic}

The overall workflow of program instrumentation is to traverse all methods in the original program and insert necessary assertions, including:
\begin{itemize}
  \item a string having a language type must be a sentence of the corresponding grammar;
  \item a value having a refinement type must satisfy the refinement condition;
  \item when calling a method, its pre-conditions must hold for the input arguments;
  \item when a method returns, its return value must fulfill its post-conditions. 
\end{itemize}

To formally present the instrumentation process, we build rules that transform an original program into an instrumented version.
At the top level, the whole program transformation is done by replacing each original method
$$
  \cMethod m(x_1: t_1, \ldots, x_k: t_k): t~\Phi~\{S\},
$$
with a new version
$$
  \cMethod m(x_1: t_1, \ldots, x_k: t_k): t~\{S'\}.
$$
The most important step is to transform the method body $S$ into $S'$ to contain all necessary assertions listed above (will be discussed soon).
And this includes checks for pre- and post-conditions, hence the contract $\Phi$ is no longer needed in the new version.

We formally present the transformation of method bodies as a judgment of the form $\Gamma \mid m \vdash S \transTo S'$, stating that under the typing context $\Gamma$ and the method $m$, a list of statements $S$ (belonging to $m$'s body) are transformed into $S'$.
The typing context $\Gamma$ maintains a list of bindings $x : t$, indicating that the local variable $x$ has type $t$.
Initially, the typing context records the types of $m$'s input arguments:
$$
  \Gamma \triangleq \{x_i: t_i \mid 1 \le i \le k\}.
$$

All the transformation rules process the statement list $S$ sequentially.
Most of them do not refer to the current method $m$ so we elide it in the rule for simplicity (hence in form $\Gamma \vdash S \transTo S'$).

\subsubsection{Declarations}

No additional assertions are needed, but the typing context gets updated by appending (denoted by a comma) a new binding recording the local variable's type:
\begin{mathpar}
  \inference[Decl]
    {\Gamma, x: t \vdash S \transTo S'}
    {\Gamma \vdash \cVar x: t; S \transTo \cVar x: t; S'}
\end{mathpar}

\subsubsection{Assignments}

Additional assertions are needed to check if the assigned value $e$ fulfills the constraint required by the type $t$ of the target variable $x$.
We extract this constraint via type normalization, say if $\norm{t} = \{y: \tau \mid \varphi\}$, then any value of $t$ must fulfill the predicate $\lambda y. \varphi$.
We define a helper function $\checkType{t}$ to obtain this predicate.
To avoid evaluating $e$ twice, we decide, in the transformed body, to first perform the original assignment and then assert that $x$ must fulfill that predicate:
$$
\inference[Assign]
  {(x: t) \in \Gamma & \checkType{t} = p & \Gamma \vdash S \transTo S'}
  {\Gamma \vdash x = e; S \transTo x = e; \cAssert p(x); S'}
$$

The added assertion is indeed redundant if $p(x)$ is simply $\cTrue$, which happens if either $t$ is a simple type or $t$ is a refinement type with a trivial $\cTrue$ condition.
Our implementation did the optimization to drop such redundant assertions.
But for the sake of generality, the transformation rules (including the subsequent) shown in the paper will keep them.

\subsubsection{Method Calls}

To handle method-related statements, we rely on several meta-properties for methods, which we precomputed for each method $m$ defined in the program:
\begin{itemize}
  \item $\sig{m}$: the type signature of the method $m$, which consists of parameter names with types, and the return type;
  \item $\pre{m}$: the pre-condition of the method $m$;
  \item $\post{m}$: the post-condition of the method $m$.
\end{itemize}
In case multiple pre-/post-conditions are declared, $\pre{m}$ and $\post{m}$ give the conjunction.
Note that $\pre{m}$ is a predicate over $m$'s input arguments, and $\post{m}$ a predicate over $m$'s input arguments and its return value.

At call sites, we need assertions:
(1) to ensure each argument fulfills the constraint required by the corresponding parameter type of the callee, and
(2) to ensure the pre-condition.

Our transformation rule is the following:
$$
\inference[Call]
  {\sig{m} = (x_1: t_1, \ldots, x_k: t_k) \to t & \pre{m} = p_m \\
   \forall i: \fresh{z_i} & \forall i: \checkType{t_i} = p_i & \Gamma, y: t \vdash S \transTo S'
  }
  {\Gamma \vdash \cVar y = \cCall m(e_1, \ldots, e_k); S \transTo
   \begin{aligned}[t]
     &\{\cVar z_i = e_i; \cAssert p_i(z_i);\}_{i=1}^{k} \\
     &\cAssert p_m(z_1, \ldots, z_k); \cVar y = \cCall m(z_1, \ldots, z_k); S'
   \end{aligned}
  }
$$
To avoid evaluating the actual arguments $e_i$ more than once, we introduce fresh variables $z_i$ to save their values.
To achieve (1), we query the signature of the callee $m$, extract the required constraints by $m$'s input types, and insert necessary assertions as done in rule Assign.
To achieve (2), we query the pre-condition of $m$ and insert an assertion to ensure the pre-condition holds for $z_i$s.
Finally, we call $m$ with $z_i$s and assign the return value to a newly declared variable $y$, appending the binding of $y: t$ to $\Gamma$, where $t$ is the return type of $m$.

\subsubsection{Return Statements}

Dual to method calls, at returning points we need assertions:
(1) to ensure the return value fulfills the constraint required by the return type, and
(2) to ensure the post-condition.

Our transformation rule is the following:
$$
\inference[Return]
      {\sig{m} = (x_1: t_1, \ldots, x_k: t_k) \to t & \post{m} = p_m &
       \fresh{z} & \checkType{t} = p}
      {\Gamma \mid m \vdash \cReturn e; S\transTo \cVar z = e; \cAssert p(z);
                     \cAssert p_m(x_1, \ldots, x_k, z); \cReturn z;}
$$
Again, to avoid evaluating the return value $e$ more than once, we introduce a fresh variable $z$ to hold its value.
To achieve (1), we query the signature of the current method $m$, extract the required constraints by $m$'s return type, and insert necessary assertions as done in rule Assign.
To achieve (2), we query the post-condition of $m$ and insert an assertion to ensure the post-condition holds for this method's input arguments $x_i$s and the return value $z$.
Because any statement after the return statement is unreachable, we do not process the rest statements $S$ as other rules do.

\subsubsection{Other Statements}

No additional assertions are needed for other statements.
We simply keep their structure and recursively process their nested bodies (if any):
\begin{mathpar}
  \inference[Assert]
    {\Gamma \vdash S \transTo S'}
    {\Gamma \vdash \cAssert e; S \transTo \cAssert e; S'}
  \and \inference[If]
    {\Gamma \vdash S_1 \transTo S_1' & \Gamma \vdash S_2 \transTo S_2' &
     \Gamma \vdash S \transTo S'}
    {\Gamma \vdash \cIf e~\{S_1\} \cElse \{S_2\}; S \transTo \cIf e~\{S_1'\} \cElse \{S_2'\}; S'}
  \and \inference[While]
    {\Gamma \vdash S_1 \transTo S_1' & \Gamma \vdash S \transTo S'}
    {\Gamma \vdash \cWhile e~\{S_1\}; S \transTo \cWhile e~\{S_1'\}; S'}
\end{mathpar}

\section{From FLAT-CORE to FLAT-X}\label{sec:migrate}

The idea behind FLAT-CORE is language agnostic and can be applied to existing mainstream programming languages.
In this section, we will discuss the general principles of migrating FLAT-CORE to a new framework FLAT-X for a choice of language X.

The most important feature of FLAT-CORE is its refinement type system with language types.
Extending the type system of the host language with a full-fledged refinement type is usually difficult:
one has to modify the compiler to check those new types, and also to be careful about the interplay of the new types with the existing ones.
A lightweight approach is, to define the language and refinement types as customized data types, without changing the type system of the host language.
For example, in an object-oriented language, one can define a new class \py{FlatType}, and define language and refinement types as its subclasses.

The next problem is how to annotate method parameters and local variables with such types, together with pre- and post-conditions for methods.
A general solution is to use a form of syntactic metadata that can be attached to the source code but ignored by the compiler/interpreter.
Many languages like Java support an annotation system that allows users to annotate classes, methods, variables, etc. with annotations created by developers.
Thus, one may build special annotations for attaching types (or simply the refinements, regarding the original types as base types), pre-, and post-conditions.
Python, on the other hand, does not have such an annotation system but supports type hints, which allows arbitrary Python expressions.
Python also supports decorators, with which one can build special annotations for pre- and post-conditions.

Once the annotations are ready, we must read them and perform program transformation.
How to perform program transformation depends on how the type annotations are given.
They are usually encoded in an \emph{abstract syntax tree} (AST) and can be read by language APIs.
For example, Python annotations are stored in AST and can be manipulated using the \py{ast} module;
Java annotations can be processed by compiler plug-ins.

Finally, library functions supported by FLAT-CORE, in particular the XPath-related ones, should also be implemented in the host language as well.
Apart from them, functions available from either the official or third-party libraries can be used to increase the expressiveness of the specifications.

\section{FLAT-PY: FLAT for Python}\label{sec:py}

Following the general principles discussed in \cref{sec:migrate}, we now present FLAT-PY, a pratical testing framework for Python.
It offers annotations (\cref{sec:py:annot}) for defining language types and refinement types, and specifying contracts for functions.
It comes with built-in types for email address, URLs, JSON, etc.
It provides a convenient \py{fuzz} function (\cref{sec:py:fuzz}) to enable random testing with the aid of language-based fuzzers.
From these annotations, we adopt a similar approach (as in \cref{sec:tyck:dynamic}) to instrument necessary type assertions via Python AST transformation (\cref{sec:py:trans}).

\subsection{Annotations}\label{sec:py:annot}

We introduce two groups of annotations for Python:
\begin{itemize}
  \item type constructors for defining new language types and refinement types, and
  \item method decorators for adding pre- and post-conditions.
\end{itemize}

Python's PEP 484\footnote{\url{https://peps.python.org/pep-0484}} standard offers a flexible way to attach type annotations: type can be arbitrary Python expressions.
To represent our language types and refinement types, which do not exist in Python's type system, we build two internal classes \py{LangType} and \py{RefinementType} for each of them and provide two type constructors as user APIs:
\begin{minted}{python3}
def lang(name: str, rules: str) -> LangType
def refine(base: type, predicate: Any) -> RefinementType
\end{minted}

The language type constructor \py{lang} takes two parameters: a type \py{name} and
a string representing the production rules, using the \emph{extended Backus-Naur form} (EBNF) defined in FLAT-CORE (\cref{sec:core}).
For example, the following expression creates a language type for a simple integer arithmetic expression using only addition and subtraction:
\begin{minted}{python3}
lang('IntExp', """
start: (number op)* number;
number: [0-9]+;
op: "+" | "-";
""")
\end{minted}

The refinement type constructor \py{refine} builds a refinement type $\{x : t \mid e\}$, taking two parameters:
\begin{itemize}
  \item a Python \py{type} that represents the \py{base} type $t$, which is either a Python built-in type (\py{int}, \py{bool}, or \py{str}) mapping to the FLAT-CORE built-in types ($\cInt$, $\cBool$, or $\cString$), or a language type constructed by \py{lang};
  \item a Python function \py{predicate} that packs the bound variable $x$ and its refinement $e$ into a lambda expression \py{lambda x: e}.
\end{itemize}
For example,
\begin{minted}{python3}
refine(JSON, lambda s: len(s) < 10)
\end{minted}
maps to the refinement type $\{s : \texttt{JSON} \mid \mathit{length}(s) < 10\}$.

One may use Python's type alias declaration to provide shorthands for language/refinement types and use them as the type annotations:
\begin{minted}{python3}
IntExp = lang('IntExp', """
start: (number op)* number;
number: [0-9]+;
op: "+" | "-";
""")
def calculate(exp: IntExp) -> int:
  ...
\end{minted}

For pre- and post-conditions, we provide two method decorators that both take a predicate as inputs:
\begin{minted}{python3}
def requires(predicate: Any)  # for pre-condition
def ensures(predicate: Any)   # for post-condition
\end{minted}
In \py{requires}, the \py{predicate} is over the input arguments of the method.
In \py{ensures}, the \py{predicate} is over not only the input arguments but also the return value, so that one is able to express input-output relations as oracles.
For example, the following contract reveals the functional correctness of converting a string with only digits into the corresponding integer value:
\begin{minted}{python3}
@requires(lambda s: s.isdigit())
@ensures(lambda s, n: int(s) == n)
def convert_digit(s: str) -> int:
  ...
\end{minted}

To avoid repeating the input arguments in the lambda expressions, we also accept a string as the \py{predicate}, which is a textual representation of a valid Python expression in which the method input arguments are automatically bound and the return value is bound to ``\code{return}''.
Using this style, the contract for \py{convert_digit} is rewritten as below:
\begin{minted}{python3}
@requires('s.isdigit()')
@ensures('int(s) == return')
def convert_digit(s: str) -> int:
  ...
\end{minted}
Both styles will be used interchangeably in the remainder of the paper for better readability.

\subsection{Type-Directed Test Generation}\label{sec:py:fuzz}

We offer a \py{fuzz} function that triggers the testing of a \py{target} function via \py{k} randomly-generated inputs:
\begin{minted}{python3}
def fuzz(target: Callable, k: int,
         using: Optional[dict[str, Generator]] = None)
\end{minted}
Each input consists of one random value per parameter of the \py{target} function, and the producers can be specified in the optional \py{using} argument.
We provide default producers for two kinds of parameters:
(1) if a default value is specified, say $v$, then we synthesize a constant producer that always yields the value $v$;
(2) if the parameter has a grammar-based refinement type $\{L \mid \varphi\}$, then we synthesize a producer that yields sentences of $L$ that also fulfill $\varphi$.
Otherwise, the user must provide one explicitly.
If the \py{target} function contains preconditions, then only values that fulfill the conditions will be used as test cases.

Synthesizing a constant producer is trivial:
\begin{minted}{python3}
def constant_producer_for_v():
  while True:
    yield v
\end{minted}

The interesting case is how to produce sentences of a CFG that also fulfill a semantic constraint.
Recall that we use ISLa as our backend fuzzer.
ISLa has its own specification language for expressing the semantic constraints, and \emph{not all} Python Boolean expressions are convertible to formulae in this language.
Our idea is to split the condition into two parts $\varphi = \varphi_1 \land \varphi_2$, where $\varphi_1$ is convertible to an ISLa formula $\varphi_\text{ISLa}$, and $\varphi_2$ is not.
The producer works in a generate-and-filter fashion: it first calls the ISLa solver to obtain a string, which is in $L$ and fulfills $\varphi_1$, but may not hold $\varphi_2$; this string is yielded only when $\varphi_2$ evaluates to \py{True}.

Technically, the condition split is realized by rewriting it into conjunctive normal form and converting as many of the conjuncts into ISLa formulae as possible, including:
\begin{itemize}
  \item standard arithmetic, comparison, and Boolean operations;
  \item string operations that have equivalences in SMT string theory;
  \item substring selection via our library functions with XPaths, together with forall/exists check on the selected strings.
\end{itemize}

\subsection{AST Transformation}\label{sec:py:trans}

Python offers a convenient \py{ast} module to parse, print, and manipulate AST nodes.
We perform program transformation by subclassing \py{ast.NodeTransformer}.
Except for the method call statement, all the statements of FLAT-CORE have equivalences in Python statements.
For them, the transformation rules are conceptually the same as the rules presented in \cref{sec:tyck:dynamic}.

As Python does not distinguish methods from functions as FLAT-CORE does, the method call statement is simply a call expression in Python (of type \py{ast.Call}).
The rule Call (in \cref{sec:tyck:dynamic}) does not apply.
But realizing a simple fact that if the method $m$ is called (from whatever method $m'$), the body of $m$ will be entered, we can instead move all the necessary assertions from the caller ($m'$) site to \emph{callee} ($m$) site, by replacing the actual arguments (\ie the fresh variables $z_i$s in rule Call) with the formal parameters of $m$.
For example, we instrument the following assertions (highlighted) for \py{convert_digit}:
\begin{minted}[highlightlines=2]{python3}
def convert_digit(s: str) -> int:
  assert s.isdigit()   # pre-condition
  ... # instrumented body
\end{minted}

So far, if any instrumented assertion fails, the stack trace will report the locations in the instrumented code rather than the original code.
This is inconvenient for the users to debug their code.
To fix this issue, in our implementation, we instrument additional statements that inject the original locations and raise our customized errors rather than Python's \py{AssertionError}.
Such customized errors will be caught, and a new stack trace will be built from the current stack trace, reporting the original locations extracted from the stack frames.
For precondition violation errors, the new stack trace reports the error locations of the caller.

\section{Case Studies}\label{sec:eval}

In this section, we apply FLAT-PY to real-world Python code fragments manipulating structured strings.
We inspect their code to learn what the expected behaviors are, and attach necessary FLAT annotations to them.
Based on the annotations, we execute our runtime checker, and if any type errors happens, we will study and understand the cause of failures.
We conducted three case studies from \cref{sec:eval:format} to \cref{sec:eval:sanitization}, targeting the following research questions:
\begin{itemize}
  \item RQ1: How much effort does the user put into specifying annotations?
  \item RQ2: How many type errors are detected using runtime checks?
  \item RQ3: What is the overhead of executing runtime checks?
\end{itemize}
Finally, we discuss the threats to validity in \cref{sec:eval:answers}.

All experiments were conducted on a MacBook Pro with Apple(R) M1 Max chip, \SI{32}{GB} memory, running Python version 3.11.9.

\subsection{Format Validation}\label{sec:eval:format}

It is common in web/GUI applications to validate that user-provided data matches the required format before writing into databases.
We studied three format validation functions taken from the PlatformIO Core project\footnote{\url{https://github.com/platformio/platformio-core/blob/591b377e4a4f7219b95531e447aec8d28fd41a79/platformio/account/validate.py}}, a platform for embedded software development.
The functions validate user names, passwords, and team names.

As an example, let us first consider the \py{validate_teamname} function:
\begin{minted}[linenos]{python3}
def validate_teamname(value):
  value = str(value).strip() if value else None
  if not value or not re.match(
    r"^[a-z\d](?:[a-z\d]|[\-_ ](?=[a-z\d])){0,19}$", value, flags=re.I
  ):
    raise BadParameter(
      "Invalid team name format. "
      "Team name must only contain alphanumeric characters, "
      "single hyphens, underscores, spaces. It can not "
      "begin or end with a hyphen or an underscore and must"
      " not be longer than 20 characters."
    )
  return value
\end{minted}

The error message at Lines 7-11 describes the format requirements for a valid team name.
We study if all strings that meet the required format can be accepted by this validation function, via random test generation.
We will adopt and compare two approaches: one with FLAT-PY based on user annotations, and the other without FLAT-PY based on a grammar-based fuzzer like ISLa alone.

\paragraph{Testing with FLAT-PY}

A valid team name only contains alphanumeric characters, single hyphens (\py{'-'}), underscopes (\py{'_'}), and whitespaces (\py{' '}), and is no longer than 20 characters.
We define a language type \py{TeamNameFormat} to reflect the above requirements:
\begin{minted}{python3}
TeamNameFormat = lang('TeamNameFormat', """
start: char{1,20};
char: [a-zA-Z0-9-_ ];
""")
\end{minted}
We use this type to restrict the input values of the validator function, using PEP 484 type hint syntax:
\begin{minted}{python3}
def validate_teamname(value: TeamNameFormat):
  ...
\end{minted}
We enable random testing of the validator function via the built-in \py{fuzz} function and we set 1,000 as the number of random test inputs:
\begin{minted}{python3}
fuzz(validate_teamname, 1000)
\end{minted}

We have provided enough annotations for FLAT-PY to set up fuzz testing.
FLAT-PY's internal program instrumentor takes care of synthesizing a proper input producer from the type signature of \py{validate_teamname}.
But due to a missing constraint that is not considered by the language type \py{TeamNameFormat}, it is fine to see the validator crash with the \py{BadParameter} error.

To attach the missing constraint ``it cannot begin or end with a hyphen or an underscore'' into the \py{TeamNameFormat} type, we define a new refinement type \py{TeamName}:
\begin{minted}{python3}
TeamName = refine(TeamNameFormat,
  lambda s: not s.startswith('-') and not s.endswith('-') and \
            not s.startswith('_') and not s.endswith('_'))
\end{minted}
This type now expresses the \emph{exact} set of valid team names, so that once we update the type signature of the validator function, no more crashes are expected:
\begin{minted}{python3}
def validate_teamname(value: TeamName):
  ...
\end{minted}

However, this time, FLAT-PY did find failing inputs such as \py{'R-_b'}.
In this example, as the special characters (hyphen and underscore) are neither the first nor the last character, it indeed follows the desired format, but is rejected by the mysterious regex.
We studied this regex carefully and realized that it encoded another constraint not mentioned in the error message: the character after the special character must be a normal alphanumeric character, as required by the lookahead assertion \py{'(?=[a-z\d])'} after \py{'[\-_ ]'}.
Thus, \py{'R-_b'} is rejected by the regex because \py{'_'} cannot occur after \py{'-'}.

\paragraph{Testing with ISLa}

One may realize the same functionality without our FLAT-PY tool, for example, to write testing code that first produces random inputs with the aid of a fuzzer like ISLa, and then sends them to the target function.

To use the ISLa solver for test generation, we need to provide a grammar with optional ISLa constraints.
For better efficiency, we define this grammar in a different manner:
\begin{minted}{python3}
import string
ALPHA_NUM = reduce(lambda c1, c2: f'"{c1}" | "{c2}"', 
                   string.ascii_letters + string.digits)
grammar = f"""
  <start> ::= <normal> | <normal><chars><normal>
  <chars> ::= "" | <char><chars>
  <normal> ::= {ALPHA_NUM} | " "
  <char> ::= <normal> | "-" | "_"
  """
constraint = 'forall <chars> s in start: str.len(s) <= 18'
\end{minted}
As required by ISLa, the \py{grammar} is written in \emph{Backus-Naur Form} (BNF), where terminals are double-quoted (\eg \py{"-"}) and nonterminals are angle-bracketed (\eg \code{<normal>}).
The constraint ``a term name cannot begin or end with a hyphen or an underscore'' is reflected in the grammar: in the \code{<start>} rule, the first and the last character of the sentence are both \code{<normal>} characters---alphanumeric (defined as \py{ALPHA_NUM}) or \py{' '}.
For any other \code{<char>} in the middle, special characters \py{'-'} and \py{'_'} are allowed.
To express there are at most 18 such middle \code{<char>}s (so that the entire length is no longer than 20), we define a recursive rule \code{<chars>} and attach an additional ISLa \py{constraint} for bounding the length.
From this \py{grammar} and \py{constraint} we instantiate an ISLa solver:
\begin{minted}{python3}
from isla.solver import ISLaSolver
solver = ISLaSolver(grammar, formula=constraint)
\end{minted}

Using this \py{solver}, we write testing code that feeds 1,000 random inputs (obtained by the \py{solve} method) to the validator function, collecting all \py{failing_inputs}:
\begin{minted}{python3}
failing_inputs = []
for _ in range(1000):
  inp = str(solver.solve())
  try:
    validate_teamname(inp)
  except BadParameter:
    failing_inputs.append(inp)
\end{minted}

\paragraph{Comparison}

FLAT-PY saves developer's time and effort in the following aspects:
\begin{itemize}
  \item EBNF-style grammars make it easier and more convenient to specify syntaxes, especially with repetitions and loops;
  \item FLAT-PY directly accepts Python expressions as constraints, thus no need to learn a new specification language such as the ISLa constraint language;
  \item No need to write code for setting up fuzzers and collecting testing results, as FLAT-PY will synthesize them automatically.
\end{itemize}

FLAT-PY also increases code readability: user-provided type annotations (\ie the grammars and constraints) serve as formal documents expressing the expected behaviors.

\paragraph{Results}

\begin{table}[t]
  \caption{Summary of testing format validation functions.
  LoA: lines of annotations.
  LoC: lines of code.
  $T_P$: time of producing random inputs (seconds).
  $T_C$: time of executing the code with instrumented checks (seconds).
  $T_O$: time of executing the original code without checks (seconds).
  }
  \label{tb:results-format}
  \begin{tabular}{llrrrr}
    \toprule
    Function & LoA/LoC & Passed/Total & $T_P$ (s) & $T_C$ (s) & $T_O$ (s) \\
    \midrule
    \py{validate_username} & 6/13 & 993/1000 & 32.857 & 64.514 & 0.001 \\
    \py{validate_password} & 5/9  & 1000/1000 & 9.457 & 8.406 & 0.001 \\
    \py{validate_teamname} & 8/13 & 993/1000 & 12.043 & 19.894 & 0.001 \\
    \bottomrule
  \end{tabular}
\end{table}

Likewise, we studied the other two format validation functions \py{validate_username} and \py{validate_password}.
\cref{tb:results-format} summarizes the results.

For RQ1, one may measure the annotation burden by the ratio of LoA to LoC.
On the three functions we studied, the ratio is less than 1.0.
However, it may exceed 1.0 and vary depending on the code and what we wish to test.
FLAT-PY tries to reduce user burden by offering convenient EBNF-style notations and accepting Python expressions directly as semantic constraints.
And for this case, the required annotations are clear from the code comments.
To this extent, we find attaching annotations is usually intuitive and pleasant.

For RQ2, FLAT-PY identified 14 failing inputs on two validation functions, from a total of 3,000 random inputs.
These random inputs were produced in \SI{54.357}{\second}, with an average speed of 55.2 inputs per second.
FLAT-PY is capable of catching logical bugs in real code, with the aid of language-based test generation.

For RQ3, from the last two columns ($T_C$ and $T_O$), we see that the instrumented checks bring significant overhead ($T_C - T_O$), increasing the time from \SI{0.01}{\second} to a few seconds and even a minute.
This is due to the time cost of parsing.
However, if the code has been well-tested against a large number of test cases and no runtime type errors were detected, the developer would have high confidence about the correctness and thus eliminate the runtime checks in production, where the overhead will no longer exist.

\subsection{File Path Safety}\label{sec:eval:safe-path}

Operating systems provide system APIs to process files, and file paths are usually represented as strings.
Safety checking is needed to prohibit security issues, for instance, unexpectedly deleting a file located in a system folder.

The following function\footnote{\url{https://github.com/macazaga/gpt-autopilot/blob/b782a1c7005eadb3f93f4dbd1f56cdf499ed265b/helpers.py\#L15}}
rejects any access to an ``unsafe'' file that is outside the trusted \code{code/} folder, hence to avoid the situation where a system file is accidentally deleted:
\begin{minted}[linenos]{python3}
def safepath(path: str) -> str:
  base = os.path.abspath("code")
  file = os.path.abspath(os.path.join(base, path))
  if os.path.commonpath([base, file]) != base:
    print(f"ERROR: Tried to access file '{file}' outside of code/ folder!")
    sys.exit(1)
  return path
\end{minted}

The expected behavior of \py{safepath} is that: if the relative \py{path} (to the base path \code{code/}) is indeed unsafe, then the function exits (and raises a \py{SystemExit} exception), otherwise it returns the \py{path} itself.
FLAT-PY offers a special decorator \py{raise_if(exc_type, predicate)} to check that if the \py{predicate} holds for the input arguments, then the function must raise an exception of type \py{exc_type}.
We use \py{raise_if} and \py{ensures} to attach post-conditions:
\begin{minted}{python3}
@raise_if(SystemExit, lambda path: not is_safe(path))
@ensures(lambda path, ret: ret == path)
def safepath(path: RelPath) -> RelPath:
  ...
\end{minted}
And we wish to fuzz it as we did in \cref{sec:eval:format}:
\begin{minted}{python3}
fuzz(safepath, 1000)
\end{minted}

The remaining problems are how to define a grammar for relative paths and a predicate that decides if a path is safe or not.
Starting at the base folder \code{code/}, a relative path can be constructed via a sequence of one of the folder changing operations:
\begin{itemize}
  \item entering a new folder: as the choice of the folder name is very unlikely to affect the behavior of the testing function, we simply pick a representative name \py{"foo"} here;
  \item going back to the parent folder via \py{".."}: this is where one can move outside the \code{code/} folder; or
  \item staying at the current folder via \py{"."}.
\end{itemize}
Thus, we introduce a new language type \py{RelPath} for relative paths as follows:
\begin{minted}{python3}
RelPath = lang('RelPath', """
start: (part "/")*;
part: "foo" | ".." | ".";
""")
\end{minted}

We observe that once we move outside of the base path, we will not be able to return again, and thus the path is unsafe.
To keep track of where we are, we use a counter \py{level} with initial value \py{0}, and:
\begin{itemize}
  \item increase it by one if entering a new folder;
  \item decrease it by one if going back to the parent folder; or
  \item do nothing if staying at the current folder.
\end{itemize}
If \py{level < 0}, we immediately know the path is unsafe.
We implement the procedure above as a safety decision function in Python:
\begin{minted}{python3}
def is_safe(path: RelPath) -> bool:
  level = 0
  for part in select_all(xpath(RelPath, '..part'), path):
    match part:
      case 'foo':
        level += 1
      case '..':
        level -= 1
        if level < 0:
          return False

  return True
\end{minted}
Note that we use our built-in function \py{select_all} and an XPath \py{'..part'} to extract the parts from parse trees.

\paragraph{Results}

For RQ1, we added (or edited) 19 lines of the annotations to the original 7 lines of code, where the ratio of LoA to LoC (= 2.7) exceeds 1.0.
This is mostly due to a nontrivial semantic predicate \py{is_safe}, but itself is not hard to specify in Python once we have the observation.
For RQ2, FLAT-PY did not find any failing inputs.
For RQ3, we measured the three time metrics as defined in \cref{tb:results-format}:
$T_P=\SI{0.383}{\second}$,
$T_C=\SI{0.340}{\second}$, and
$T_O=\SI{0.006}{\second}$.
Again, the overhead was significant.

\paragraph{Remark}

Compared with the refinement predicate of \py{TeamName} defined in \cref{sec:eval:format}, the predicate \py{is_safe} here is more complex and hence may not be easily defined in other specification languages such as ISLa.
In particular, the traversal of the parts, which, if defined in a functional specification language, typically requires higher-order functions that not all of them support.

\subsection{File Path Sanitization}\label{sec:eval:sanitization}

Another situation where unsafe file paths cause security issues is the existence of certain nonstandard, dangerous characters, which may cause damage to the file system or lead to Shell command injection.
We study a function\footnote{\url{https://github.com/joeken/Parenchym/blob/63864cdaff76b9aa1b8dbe795eb537b5be5add3a/pym/lib.py\#L374}}
that sanitizes a potentially unsafe path into a safe version, free of certain dangerous characters (see docstring for details):
\begin{minted}{python3}
def safepath(path: str) -> str:
  """
  Returns safe version of path.
  Safe means, path is normalised with func:`normpath`, and all parts are
  sanitised like this:
  - cannot start with dash ``-``
  - cannot start with dot ``.``
  - cannot have control characters: 0x01..0x1F (0..31) and 0x7F (127)
  - cannot contain null byte 0x00
  - cannot start or end with whitespace
  - cannot contain '/', '\', ':' (slash, backslash, colon)
  - consecutive whitespaces are folded to one blank
  """
  ...  # 11 lines of code elided
\end{minted}

Our goal is to test if \py{safepath} can always returns a sanitised path for any potentially dangerous path, so we need to define new language types \py{Path} and \py{SanitizedPath} for each of them, and refine the type signature of \py{safepath}:
\begin{minted}{python3}
def safepath(path: Path) -> SanitizedPath:
  ...
\end{minted}
Again, we will apply the fuzzing technique for detecting bugs:
\begin{minted}{python3}
fuzz(safepath, 1000)
\end{minted}

\paragraph{Input grammar}

To define a grammar for \py{Path}, let us extend the grammar for \py{RelPath} defined in \cref{sec:eval:safe-path} with absolute paths and more general file names with ASCII characters:
\begin{minted}{python3}
Path = lang('Path', """
start: "/" | "/"? part ("/" part)*;
part: (%x0-2E | %x30-7F)*;
""")
\end{minted}
As the parent folder \py{'..'} and the current folder \py{'.'} are both just ASCII strings, each part is simply an ASCII character sequence without the Unix separator \py{'/'} (of code point \py{0x2F}).
Here we encode this using a special clause supported by FLAT-PY of the form ``\verb|%x|$k_1$\verb|-|$k_2$'', inspired by RFC 5234 (Augmented Backus–Naur form), to mean any character in between the hexadecimal code points $k_1$ and $k_2$ (both inclusive).

\paragraph{Output grammar}

A \py{SanitizedPath} is still a \py{Path} but without certain characters (at certain places) as detailed in the docstring of \py{safepath}.
We define \py{SanitizedPath} as a refinement type based on \py{Path}:
\begin{minted}{python3}
SanitizedPath = refine(Path, is_sanitized)
\end{minted}
where the refinement predicate is defined as follows:
\begin{minted}{python3}
def is_sanitized(path: Path) -> bool:
  parts = select_all(xpath(Path, '..part'), path)
  return not any(
    [part.startswith('-') or part.startswith('.') or part.startswith(' ') or
     part.endswith(' ') or
     any([is_not_allowed(ch) for ch in part]) or
     (' ' * 2) in part
     for part in parts])
\end{minted}
It checks if no part violates the constraints listed in the docstring:
\begin{itemize}
  \item starts with dash (\py{'-'}), dot (\py{'.'}), or whitespace (\py{' '});
  \item ends with whitespace;
  \item contains any control characters, null byte, slash (\py{'/'}), backslash (\py{'\\'}), and colon (\py{':'}), as defined in an auxiliary predicate \py{is_not_allowed} (see below);
  \item contains consecutive whitespaces, \ie at least two whitespaces (\py{' ' * 2}).
\end{itemize}

The auxiliary predicate \py{is_not_allowed} is defined as follows:
\begin{minted}{python3}
def is_not_allowed(char: str) -> bool:
  return ord(char) in range(0, 31 + 1) or ord(char) == 127 or
         char in {'/', '\\', ':'}
\end{minted}

With the input and output grammars, we are ready to fuzz \py{safepath}.
The fuzzer did not produce any failing input.
However, we found this testing inadequate due to the potentially low coverage of ``unusual'' inputs containing the not-allowed characters, as the fuzzer would instead choose a normal character (like alphanumeric characters) with a higher probability.

\paragraph{Input grammar, revised}

To increase the frequency of such unusual inputs, which are more likely to uncover bugs, we \emph{refine} \py{Path} as a subtype \py{UnusualPath}:
\begin{minted}{python3}
UnusualPath = lang('UnusualPath', """
start: "/" | "/"? part ("/" part)*;
part: char*;
char: [abAB0] | [-.\\: ] | %x0-4;
""")
\end{minted}
In contrast to \py{Path}, this \py{UnusualPath} type only considers a subset of \py{char}s, which are divided into three groups, each containing a small but the same number of (five) characters as representatives, and hence is chosen by the fuzzer with an equal probability:
\begin{itemize}
  \item normal alphanumeric characters \py{{'a', 'b', 'A', 'B', '0'}},
  \item control characters from code point 0x0 to 0x5 (in Python syntax, they are represented as \py{'\x01'} to \py{'\x04'}), and
  \item special characters \py{{'-', '.', '\\', ':', ' '}} that are not allowed anywhere.
\end{itemize}

Without changing the type signature of \py{safepath}, \ie a function from \py{Path} to \py{SanitizedPath}, we tell the \py{fuzz} function to generate random paths using the above \py{UnusualPath} by specifying the optional argument \py{using} (where \py{lang_generator} convert a language type to a \py{Generator}):
\begin{minted}{python3}
fuzz(safepath, 1000, using={'path': lang_generator(UnusualPath)})
\end{minted}
This time, we identified failing inputs (will be shown soon).

\paragraph{Results}

\begin{table}[t]
  \caption{Failure test cases of \py{safepath}.}
  \label{tab:failure-safepath}
  \begin{tabular}{llllll}
  \toprule
  \# & Input \py{path}     & Length  & Output    & Constraint violated \\
  \midrule
  1  & \py{'/\x03:..a'}                     & 6       & \py{'..a'}      & start with \py{'.'}   \\
  2  & \py{'/\x00\x040//:-/\x00\\\x01/'}    & 13      & \py{'0/-'}      & start with \py{'-'}   \\
  3  & \py{':-/\\B0/./\x03B\x02\x00\x04/'}  & 15      & \py{'-/B0/B'}   & start with \py{'-'}   \\
  4  & \py{'//:.\x03\x01\x030//b'}          & 11      & \py{'.0/b'}     & start with \py{'.'}   \\
  5  & \py{'/:\\-'}                         & 4       & \py{'-'}        & start with \py{'-'}   \\
  6  & \py{'/-.:-\\\\'}                     & 7       & \py{'-'}        & start with \py{'-'}   \\ 
  7  & \py{':-\x03:'}                       & 4       & \py{'-'}        & start with \py{'-'}   \\ 
  8  & \py{'\\-:'}                          & 3       & \py{'-'}        & start with \py{'-'}   \\ 
  \bottomrule
  \end{tabular}
\end{table}

For RQ1, we added (or edited) 23 lines of the annotations to the original 31 lines of code (\ie LoA/LoC = 0.75).
Like in \cref{sec:eval:format}, the required annotations are clear from code comments so we performed a straightforward translation.

For RQ2, FLAT-PY found 8 failing inputs, with lengths 3-15.
The information of these failure test cases is summarized in \cref{tab:failure-safepath}.
For each test case, the table lists the input argument \py{path} (we fix the argument \py{sep} to be the Unix path separator \py{'/'}) with its length, the output, and the violated condition that explains why the output is not sanitized.
The common failure reason is that the outputs start with a disallowed character (here, \py{'-'} or \py{'.'}).

For RQ3, we measured three time metrics as defined in \cref{tb:results-format}:
$T_P=\SI{1.071}{\second}$,
$T_C=\SI{4.755}{\second}$, and
$T_O=\SI{0.001}{\second}$.
Again, the overhead was significant.

\paragraph{Remark}

To increase the probability of hitting bugs, we may have to increase the frequency of extreme/corner cases in the generated random inputs, which can be expressed in a new grammar adapted for fuzzing.
But we should not use it in the type signature of the function as it is incomplete---it only contains a subset of all valid inputs.
Instead, FLAT-PY allows one to customize the producers in the \py{fuzz} function, so that the original type signature is kept.

\subsection{Threats to Validity}\label{sec:eval:answers}

External validity is the major threat in our case studies: there are only three cases and we are unsure if the same results can be generalized to other codebases, in particular the annotation burden may vary.
When the tested code becomes more complex, one may put more effort on tuning the annotations (as in \cref{sec:eval:sanitization}) to uncover more interesting bugs.
However, one does not have to do so at the very beginning but refine the annotations \emph{iteratively}: start with attaching grammars merely, check for potential format violations, fix them, and refine the annotations to gradually consider more and more semantic correctness.
In this way, the annotation effort will be under control.

\section{Related Work}\label{sec:related}

\paragraph{Safe strings}

It is a widespread problem that many strings have latent structure but type checkers of mainstream programming languages cannot reach it.
The TypeScript community introduces \emph{regex-validated strings} to guarantee that a string value must match a given regex.
However, the memory of that structure is discarded after validation and thus cannot be further reused.
To address this issue, \citet{SafeStrings} present \emph{SafeString}, a programming model where one specifies the latent structure as a grammar, and once a string is validated using a parser, its underlying structure is stored as an object in the memory so that rechecking is saved when updates come.

Both SafeString and our FLAT aim to improve the type safety of strings by exposing their latent structures.
The major difference lies in the level of automation.
SafeString defines an interface for safe strings.
But it is the programmer's responsibility to realize a particular kind of safe string: to specify its structure, to build its parser, and to define a group of type-safe operations if needed.
Therefore, it fits better when developing new software and refactoring existing codebases.
On the other hand, FLAT is an extension to the type system of an existing programming language, offering users annotations to attach types and specifications.
Apart from annotations, no additional code is mandatory in FLAT: parsers are automatically derived from grammars, and standard string operations apply to language types.
Type checking is automated.
To this end, FLAT applies to legacy codebases as well.

June \cite{June} is a follow-up work of SafeString.
It exposes the latent structure of strings to test generation tools via annotation-driven code transformation and equips them with built-in safe string annotations for delimited strings, file paths, emails, dates, etc, each of which is implemented as a safe string definition by subclassing the core \py{SafeString} class.
Compared with SafeString, June is burdenless if using the built-in safe strings merely.
When it comes to other kinds of safe strings, it inherits the inconvenience of SafeString too--- the user has to subclass \py{SafeString} to realize the custom definition.
In contrast, FLAT provides a principled type system where one defines new types in a declarative and simple way.

\paragraph{Refinement types}

The concept of \emph{refinement types} was first invented in ML's type systems \cite{Refinement-Types}.
The refinement type system allows more errors to be exposed at compile time, by refining user-defined algebraic data types in standard ML.
For instance, a singleton list as a subtype of list.

A more modern style of refinement types is \emph{Logically Qualified Data Types} \cite{Liquid-Types}, abbreviated to Liquid Types.
Liquid types allow programmers to enjoy many of the benefits of dependent types with minor annotation effort, by encoding refinements as first-order logic predicates, with standard operations on common data types such as integers, bit vectors, and arrays.
This idea is then introduced in many other programming languages, like Haskell \cite{Liquid-Haskell}, TypeScript \cite{Refinement-TypeScript}, and Java \cite{Liquid-Java}.
However, none of them have focused on how to refine the string type to inject syntactic and semantic requirements of strings, as FLAT does.

\paragraph{Primitive obsession}

There are conceptually different types like file paths and UUIDs but are represented into a single built-in type ``string''.
Such a code smell is called ``primitive obsession'' \cite{Primitive-Obsession-1,Primitive-Obsession-2} and is considered a bad practice to use one primitive type for values with conceptually different types.

One possible solution is to define new types that are distinguishable by the type system.
For instance, in an object-oriented language, one may use two separate class types \py{Path} and \py{UUID}, instead of the primitive string type, to distinguish file paths from UUIDs.
This solution brings abstraction overhead (both for developers and runtime): one needs to wrap up a string into a \py{Path} object and when processing, unwrap the string value.
Scala 3 introduces \emph{opaque types} \cite{Scala-Opaque} for defining aliases as conceptually different types from a primitive type.
This distinction is made only at the type-system level for type safety but not at runtime for efficiency.

The above idea of distinguishing types using names is more like a \emph{nominal} approach,
while FLAT, which adopts a refinement type system, is a \emph{structural} approach---the refinement type itself conveys structural information about the valid value set.

\paragraph{Language-based testing}

Test generation has been studied for a long to improve test coverage and reduce human labor by automatically producing test inputs.
To fit programs accepting structural text, such as compilers and interpreters, syntax-aware, grammar-based input generation has been studied.
\citet{Grammar-Whitebox} present \emph{grammar-based white-box fuzzing}, which increases the test coverage on Internet Explorer 7 JavaScript interpreter.
CSmith \cite{CSmith} generates random C programs from a subset of the C grammar with hand-crafted generators, which detected hundreds of correctness bugs in mainstream C compilers like GCC and LLVM.
LangFuzz \cite{LangFuzz} enables \emph{black-box fuzzing} based on a CFG, and discovers security issues on the Mozilla JavaScript engine.

The above grammar-based test generators can produce syntactically valid inputs, but sometimes semantics validity is also important.
Input algebras \cite{Input-Algebras} allows one to express semantic requirements as a Boolean combination of patterns, serving as a specialization of the original CFG.
A grammar transformer computes the specialized grammar, from which any standard grammar-based fuzzer can produce inputs conforming to the requirements.
ISLa \cite{ISLa}, on the other hand, natively supports semantic constraints expressed in its specification language, which is essentially a first-order logic over derivation trees of strings.
These language-based, semantic-aware fuzzers enable type-directed testing in FLAT: types are translated into a CFG with logical constraints that depict the valid input space, from which random inputs are produced for testing.

\paragraph{Language mining}

FLAT relies on user annotations to check correctness at runtime.
The out-of-the-box FLAT built-in types and the rich EBNF notations ease the process of adding user annotations, but if they could be automatically inferred, our approach would become even more attractive.
The key problem to realize this is language mining---automatically learning grammars from programs or examples.

\citet{Grammar-Inference-Ad-Hoc-Parsers} propose a static and automated grammar inference system for ad hoc parsers.
Their intuition is that ad hoc parsers are usually language recognizers and internally maintain state machines for parsing.
Through a collection of the constraints that the input must fulfill from the parser code, they construct a regular expression $r$ as the inferred language accepted by the parser.
Annotating this parser function with input type $r$, FLAT can test both the parser function itself and other functions referring to it.

There are also dynamic approaches that mine input grammars from a set of positive examples.
Autogram \cite{Autogram} uses \emph{dynamic tainting} to track the data flow from the input string to its fragments found during execution.
A grammar that describes the hierarchical structure of the fragments is generalized from the structure of the call tree.
However, if any fragment of the input is not stored in some variable, there will be no data flow to learn from.
Mimid \cite{Mimid} tackles this problem by tracking all accesses of individual characters in the input, regardless of their usage.
Arvada \cite{Arvada} targets a black-box setting where the program code is unavailable.
Instead, it relies on an oracle to tell if an input is valid or not.
It attempts to create the smallest CFG possible consistent with all examples, via tree bubbling and merging.

\section{Conclusion and Future Directions}\label{sec:conclusion}

String is a universal type used to encode all kinds of data that indeed have different latent structures.
To distinguish them, we propose FLAT, regarding formal languages as types.
Such language types expose the underlying syntactic structure of strings.
With semantic constraints as refinements, context-sensitivity is also taken into account.
Such an expressive type system allows us to catch bugs as early as possible, and meanwhile, to automatically generate random inputs that conform to the user's assumptions, higher the chances of hitting more interesting and deeper bugs.

We implement FLAT in Python and present a practical testing framework FLAT-PY.
We conducted case studies on real Python code fragments and detected logical bugs via random testing based on a reasonable amount of user annotations.
The source code and experimental data are publicly available:
\begin{center}
  \url{https://github.com/paulzfm/flat-py}.
\end{center}

An interesting future research direction is, based on the user-provided types and pre-/post-conditions, we can check them at compile time using static approaches from the formal verification community, so that the functional correctness is guaranteed.

\bibliographystyle{ACM-Reference-Format}
\bibliography{bib}

\end{document}